\documentclass[journal]{IEEEtran}
\IEEEoverridecommandlockouts
\usepackage{times,amsmath,color,amssymb,graphicx,epsfig,cite,psfrag,subfigure,algorithm,balance}
\usepackage{amsfonts,pifont,enumerate,cases}
\usepackage{mathrsfs} 
\usepackage[table]{xcolor} 
\usepackage{verbatim} 
\usepackage{bm}
\usepackage{cuted,stfloats}
\usepackage{algorithm}
\usepackage{algorithmic}

\usepackage{longtable}
\usepackage{blindtext}
\usepackage{multirow}
\usepackage{float}
\usepackage{threeparttable}
\usepackage{makecell}
\usepackage[utf8]{inputenc}
\usepackage{url}
\usepackage{booktabs}
\usepackage{amssymb}
\usepackage{bbding}
\usepackage{pifont}
\usepackage{wasysym}
\usepackage{utfsym}
\usepackage{fontawesome}
\usepackage[algo2e,ruled,vlined,linesnumbered,lined,boxed,commentsnumbered]{algorithm2e}
\usepackage{amsmath,mathtools}
\usepackage[
    colorlinks=true,
    linkcolor=blue,
    citecolor=blue,
    urlcolor=magenta
]{hyperref}

\begin{document}
\title{Adaptive Beam Hopping and Power Control for Dual-Layer Over-the-Air Online Federated Learning in LEO Satellite Networks}
\author{Zhendong Li, Shaojie Wang, Zhou Su, Zihao Zhang, Haixia Peng, \\Nan Cheng, Ying Wang, and Wen Chen
\thanks{Zhendong Li is with the School of Information and Communication Engineering, Xi'an Jiaotong University, Xi'an 710049, China, is also with the State Key Laboratory of Integrated Services Networks and School of Telecommunications Engineering, Xi’an 710126, China (email: lizhendong@xjtu.edu.cn). Shaojie Wang, Zihao Zhang, and Haixia Peng are with the School of Information and Communication Engineering, Xi'an Jiaotong University, Xi'an 710049, China (email: wessel@stu.xjtu.edu.cn; zhangzh8219@stu.xjtu.edu.cn; haixia.peng@xjtu.edu.cn). 

Zhou Su is with the School of Cyber Science and Engineering, Xi'an Jiaotong University, Xi'an 710049, China (email: zhousu@ieee.org). 

Nan Cheng is with the State Key Laboratory of Integrated Services Networks and School of Telecommunications Engineering, Xidian University, Xi'an 710126, China (e-mail: nancheng@xidian.edu.cn). 

Ying Wang is with the State Key Laboratory of Networking and Switching Technology, Beijing University of Posts and Telecommunications, Beijing 100876, China (e-mail: wangying@bupt.edu.cn). 

Wen Chen is with the Department of Electronic Engineering, Shanghai Jiao Tong University, Shanghai 200240, China (e-mail: wenchen@sjtu.edu.cn). 

(Corresponding author: Zhou Su)}
\vspace{-1.5em}
}
\maketitle
\thispagestyle{empty}

		\maketitle
		
\begin{abstract}
This paper investigates over-the-air (OTA) computation enabled online federated learning (FL) in low-Earth orbit (LEO) satellite networks. 
Specifically, we consider a dual-layer OTA aggregation architecture, where ground devices upload analog model updates to serving satellites via uplink OTA aggregation, and satellites forward the aggregated signals to a data processing center through the second round OTA aggregation. 
Then, we formulate a long-term data-utilization maximization problem in which devices continuously collect new data and untrained samples gradually lose freshness. 
The problem is subject to the satellite beam budget, transmit-power limit, and global mean squared error (MSE) constraint that governs end-to-end aggregation distortion. 
This yields a coupled mixed-integer nonlinear programming (MINLP) problem, involving tightly coupled discrete beam-hopping decisions and continuous power control. 
Due to the combinatorial action space and non-convex constraints, the problem is NP-hard and computationally intractable. 
Furthermore, the time-varying satellite topology and dynamic data generation render it a sequential decision-making problem, necessitating adaptive online scheduling.
To address these issues, we cast the problem as a Markov decision process and develop a proximal policy optimization (PPO)-based deep reinforcement learning framework that jointly optimizes adaptive beam hopping and power control, using an MSE-aware reward to balance data utilization and aggregation accuracy. Numerical simulation results verify that the proposed algorithm consistently outperforms other benchmark schemes, achieving superior long-term data utilization and faster FL convergence while satisfying the MSE requirement.
\end{abstract}
		
		\begin{IEEEkeywords}
			Dual-layer OTA aggregation, online federated learning, LEO satellite networks, beam hopping, PPO.
			
		\end{IEEEkeywords}
		
\section{Introduction}
\IEEEPARstart{E}{dge} intelligence has become a key enabler for data-driven services in future wireless and IoT ecosystems, where massive numbers of sensors, mobile devices, and smart objects continuously generate privacy-sensitive data at the network edge~\cite{EI_ProcIEEE}.
To avoid the prohibitive latency and bandwidth cost of uploading raw data to the cloud, federated learning (FL) has emerged as a distributed training paradigm that keeps data local while only exchanging model updates or gradients with an aggregator~\cite{FL_MEN_Survey}. 
FL has been widely considered for a broad range of edge applications such as mobile edge networks, large-scale IoT systems, and network management in 5G and beyond~\cite{fediot,Fednet}.
However, most existing FL deployments and designs implicitly rely on pervasive terrestrial connectivity and ground infrastructures to provide reliable aggregation and coordination services~\cite{WFL_DigAna,FL_Subspace_Compression}.
In remote or extreme environments such as oceans, deserts, and disaster-stricken areas, deploying dense terrestrial edge servers is not only economically prohibitive but also physically vulnerable. Consequently, ensuring reliable coverage and continuous reachability to an aggregator becomes a primary bottleneck for sustaining FL at scale. 

To overcome the coverage and reachability limitations imposed by terrestrial infrastructures, recent research has increasingly considered satellite networks as programmable computing platforms that can extend distributed intelligence to wide-area scenarios~\cite{satcomp}.
Within this context, prior studies on satellite-based computing networks highlighted that integrating FL into satellite architectures could effectively exploit on-orbit resources while preserving data locality, thereby supporting privacy-aware intelligence over globally distributed devices~\cite{satfl_wcmag}. 
The architectural design and orchestration issues of FL over large-scale constellations were discussed in~\cite{fl_sat_constellations}.
Ground-assisted aggregation was proposed to relieve the limited onboard resources of low-Earth orbit (LEO) satellites in~\cite{groundfl_leo}.
More specialized frameworks further incorporated computation offloading and decentralized coordination among satellites to improve learning efficiency under time-varying connectivity~\cite{FedLEO}.
In space--air--ground integrated networks, federated reinforcement learning was also leveraged for traffic offloading and resource allocation under heterogeneous and dynamic topologies~\cite{frl_sagin}.
In parallel, a body of work studied joint communication--computation resource management for satellite edge computing, such as two-timescale service deployment and task scheduling, and mobility-aware computation offloading for fast-moving LEO satellites~\cite{sec_twotimescale,sec_mobility}, and related extensions have addressed joint sensing, communication, and computing optimization in low-altitude aerial edge computing~\cite{lawn_jscc_security}, where performance safeguarding against channel-access attacks has also been pursued~\cite{lawn_isac_attack}.
Despite this progress, most existing studies treated model aggregation mainly at the link or network layer and relied on conventional orthogonal access schemes, leaving the impact of physical-layer aggregation mechanisms and satellite-specific beam management largely underexplored.
Specifically, satellite-FL systems still suffered from two tightly coupled bottlenecks.
One was achieving communication-efficient aggregation over power- and bandwidth-limited satellite links.
The other was agile beam-level resource allocation under fast-varying coverage and multi-beam constraints.
These bottlenecks called for learning-centric designs that revisit both aggregation and scheduling.

To address the communication-efficient aggregation challenge, over-the-air (OTA) computation, also known as AirComp, has been advocated as a physical-layer technique that exploits the waveform superposition property of wireless multiple-access channels~\cite{fedOTA}.
By enabling the aggregator to receive an analog superposition of local updates in a single channel use, OTA computation provided a scalable mechanism to reduce uplink latency in wireless FL~\cite{comudo}. 
Building on this concept, extensive research efforts have been devoted to optimizing OTA-FL from various perspectives.
In terms of fundamental design and analysis, customized analog aggregation schemes were developed to minimize aggregation error~\cite{guoAnalogAgg}, while one-bit quantization techniques were proposed to further enhance communication efficiency~\cite{zhuOneBitAgg}.
To mitigate the impact of channel noise on convergence, learning rate optimization strategies were also investigated in~\cite{xuLRaircomp}.
Beyond theoretical analysis, practical deployment issues have received significant attention.
Resource-aware scheduling policies were designed to explicitly account for device energy constraints~\cite{sunDynamicAirFL}.
Addressing the challenges of non-ideal wireless environments, robust frameworks were proposed to handle heterogeneous data distributions~\cite{seryHeteroAirFL}, imperfect synchronization~\cite{shaoMisalignedAirFL}, and multi-cell interference~\cite{wangInterferenceAirFL}.
Furthermore, recent advancements have incorporated emerging physical-layer technologies to enhance aggregation reliability.
For instance, reconfigurable intelligent surfaces (RIS) and non-orthogonal multiple access (NOMA) were leveraged in~\cite{niRISNOMAAirFL} and~\cite{zhengRISAFL} to improve coverage and interference management, respectively.
Additionally, digital OTA aggregation was explored in multi-antenna systems to achieve a better balance between accuracy and robustness~\cite{wangDigitalAirFL}.
Although these works collectively demonstrate that OTA computation can substantially alleviate the uplink bottleneck, they were developed almost exclusively for terrestrial networks with relatively stable coverage and readily coordinated resources.
Their applicability to satellite systems, characterized by high-mobility dynamics, stringent beam and power budgets, and large-scale spatial heterogeneity, remains largely unexplored.
This gap leaves open the question of how OTA aggregation can be effectively adapted to learning-centric objectives under satellite-specific constraints.

In addition, to address the resource limitations and spatial heterogeneity in satellite environments, beam hopping (BH) has emerged as a key mechanism for flexible resource management~\cite{BH}.
By dynamically illuminating different subsets of spot beams over time, BH allows satellites to match highly non-uniform and time-varying traffic demands while efficiently reusing spectrum and power resources.
To exploit these benefits, optimization-based BH designs were extensively investigated for throughput maximization and spectrum sharing.
For instance, resource allocation strategies were developed for LEO satellites in spectrum-sharing scenarios~\cite{BH}, and joint precoding schemes were proposed to support flexible coverage in high-throughput systems~\cite{BH3}.
More recently, sophisticated frameworks incorporated digital-twin models, where demand-aware BH patterns were jointly optimized with transmit power to balance traffic loads and enhance system utilization~\cite{BH2}.
In parallel, deep reinforcement learning (DRL) was introduced to address the combinatorial complexity of beam scheduling and dynamic traffic variations, e.g., multi-objective DRL algorithms were applied to dynamic BH in satellite broadband systems~\cite{BH_DRL1}, while multi-agent DRL frameworks were constructed for joint beam-pattern and bandwidth allocation~\cite{BH_DRL2}.
These works demonstrate that BH can provide highly agile and fine-grained spatial-temporal resource adaptation.

The integration of OTA-enabled FL into BH-assisted LEO networks faces distinct challenges arising from the complex interplay between physical-layer signal processing, beam scheduling, and learning performance. Most existing BH strategies prioritize communication metrics like throughput and overlook learning-centric goals such as convergence speed and aggregation accuracy. This limitation is critical in online FL scenarios, where limited satellite beams must be dynamically scheduled to capture continually arriving data with strictly bounded freshness. Moreover, the high mobility of LEO satellites and the inherent distortion of OTA aggregation invalidate terrestrial schemes, which typically rely on static datasets and stable channel conditions.
Motivated by these challenges, this paper investigates adaptive BH and power control for OTA-enabled online FL, specifically targeting the complex interplay between aggregation quality, resource scarcity, and data freshness.
Unlike most existing OTA FL schemes that focus on terrestrial cellular systems, and distinct from conventional satellite resource allocation that relies on orthogonal access and ignores learning performance, we develop a learning-centric framework.
This approach explicitly accounts for the time-varying value of data and the global mean squared error (MSE) constraint, thereby realizing the synergy of efficient aggregation, adaptive scheduling, and reliable learning in dynamic satellite networks.
The main contributions of this paper are summarized as follows:
\begin{itemize}
  \item 
  A novel dual-layer OTA-enabled online FL framework is proposed for LEO satellite networks, where ground devices continuously collect new data and upload analog model updates to serving satellites via uplink OTA aggregation, and the satellites further forward aggregated signals to a data processing center through a second OTA link. 
  The framework explicitly captures data freshness via a time-discounted data-amount model and incorporates a global mean squared error (MSE) constraint, thereby bridging physical-layer OTA aggregation and learning performance in a dynamic satellite environment.
  
  \item 
  We formulate a joint optimization problem of adaptive BH and power control subject to satellite-specific beam budgets, power limits, and aggregation error thresholds.
  Recognizing the NP-hard nature of the resulting mixed-integer nonlinear programming (MINLP) and the time-varying system topology, we recast the optimization as a sequential decision-making problem within a Markov decision process (MDP) framework. 
  To tackle the high-dimensional and continuous-discrete hybrid action space, a proximal policy optimization (PPO)-based DRL algorithm is developed to learn efficient policies directly from interaction with the OTA online FL environment.

  \item 
    Extensive simulations on two representative FL tasks demonstrate that the proposed PPO-based algorithm consistently outperforms several DRL baselines and a channel-gain-based greedy policy in terms of convergence speed, final accuracy, and long-term data utilization.
    Furthermore, detailed sensitivity analyses regarding the beam budget and global MSE threshold are provided, revealing profound insights into how these satellite-specific design parameters influence the performance of OTA-enabled online FL.
    
\end{itemize}

The remainder of this paper is organized as follows.
In Section~II, the system model for the dual-layer OTA-enabled online FL architecture is introduced, and the joint BH and power control problem is formulated.
Section~III details the proposed PPO-based DRL framework designed to solve the resulting MDP.
Subsequently, numerical simulation results are presented in Section~IV to demonstrate the superior performance of the proposed algorithm.
Finally, Section~V concludes the paper and outlines directions for future work.
\section{System Model and Problem Formulation} \label{SEC:II}
\begin{figure}[t]%
			\centering			\includegraphics[width=0.5\textwidth]{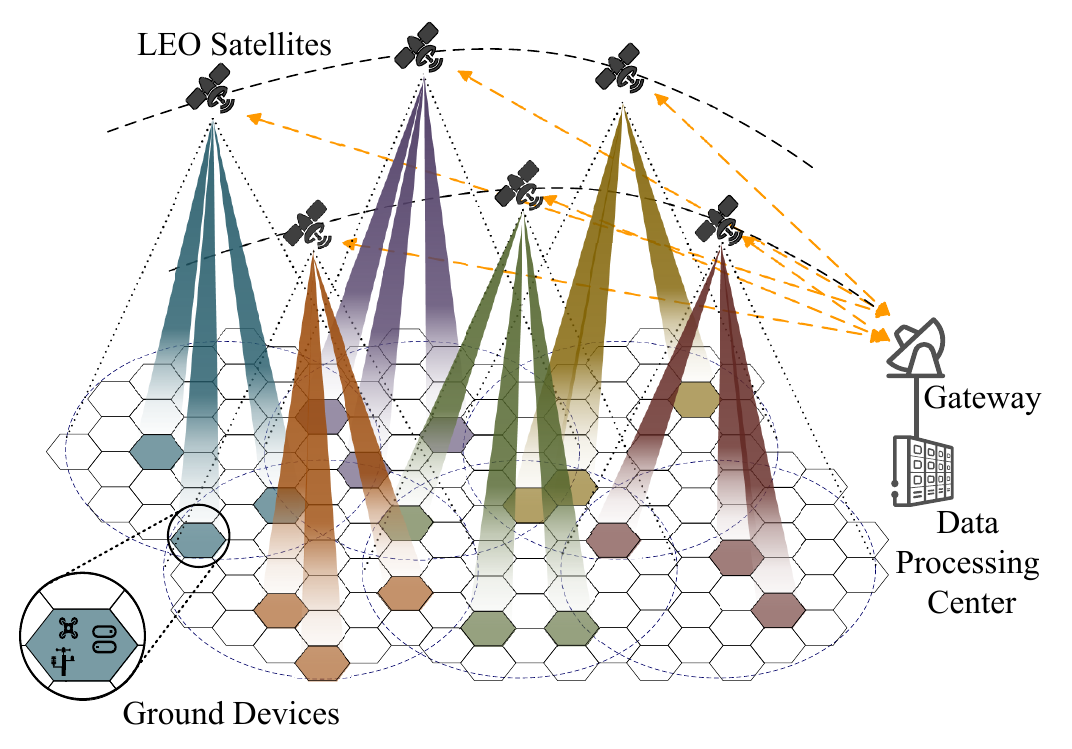}
			\caption{BH framework for OTA-enabled online FL in LEO satellite networks.}
			\label{fig:system model}
		\end{figure}

As illustrated in Fig. \ref{fig:system model}, we consider an online FL system comprising $K$ ground devices indexed by $k \in \mathcal{K}=\{1, \ldots, K\}$, a constellation of $N$ LEO satellites indexed by $n \in \mathcal{N}=\{1, \ldots, N\}$, and a ground data processing center (DPC). 
Note that the DPC operates as a logical central aggregator. In practical deployments, it is physically realized by a set of distributed ground gateways globally interconnected via high-speed optical fiber links, which guarantees continuous connectivity with the moving satellites. 
The service area is partitioned into $S$ cells indexed by $s \in \mathcal{S}=\{1, \ldots, S\}$, where the set of devices associated with cell $s$ is denoted as $\mathcal{K}_s \subseteq \mathcal{K}$. In this network, each satellite $n$ covers a dynamic set of cells $\mathcal{S}_n(t) = \{s_{n,1}(t), \ldots, s_{n,C}(t)\}$ containing up to $C$ cells, and it activates a beam set $\mathcal{V}_n(t)$ with cardinality $|\mathcal{V}_n(t)| \le V$ to serve them. Due to overlapping footprints \cite{multisat}, a cell may be simultaneously visible to multiple satellites, i.e., $\mathcal{S}_i(t) \cap \mathcal{S}_j(t) \neq \varnothing$, which provides spatial diversity for scheduling but also introduces potential inter-satellite interference.
Note that the duration of a single communication round is typically on the order of seconds, whereas the visible time of a LEO satellite over a specific cell spans several minutes. Therefore, the satellite-to-ground topology and the corresponding channel statistics can be reasonably assumed as quasi-static within each communication round. This timescale separation ensures that the connection remains stable and no inter-round beam switching is triggered during the local training and aggregation phases. 
Within this architecture, each device performs local training based on its private data, and the updated model parameters are first aggregated via OTA computation at the serving satellites. Then, the intermediate model parameters are forwarded to the DPC, where a global model is maintained and broadcast back to the network.

The FL process proceeds in $T$ global communication rounds indexed by $\mathcal{T} =\left\{ t\ \big|\ t=1,\cdots ,T \right\}$, each of duration $\tau$. 
In each round $t$, the following steps are executed:
        \begin{enumerate}
            \item The DPC broadcasts the current global model $\mathbf{z}(t)$ to all satellites $n \in \mathcal{N}$.
            
            \item Each satellite $n$ selects a subset of its currently covered cells $\mathcal{S}_n(t)$, denoted by $\mathcal{A}_n(t) \subseteq \mathcal{S}_n(t)$, satisfying $|\mathcal{A}_n(t)|=|\mathcal{V}_n(t)| \leq V$. 
            To avoid redundant aggregation and strong inter-satellite interference, no two satellites are allowed to select the same cell simultaneously, i.e., ${{\mathcal{A}}_{i}}\left( t \right)\bigcap {{\mathcal{A}}_{j}}\left( t \right)=\varnothing ,\forall i,j\in \mathcal{N}$. 
            Let $\mathcal{K}(t)$ denote the set of participating devices located in them.
      
            \item Each satellite transmits one beam to each selected cell $s \in \mathcal{A}_n(t)$, delivering $\mathbf{z}(t)$ to all devices in the cell.

            \item Each device $k \in \mathcal{K}_s$, where $s \in \mathcal{A}_n(t)$, uses its local dataset $\mathcal{D}_k(t)$ to update the global model and produce the local model $\mathbf{z}_k(t+1)$.
            
            \item Satellites perform local aggregation of updated models received from devices in their selected cells, resulting in intermediate models $\mathbf{z}_n(t+1)$.
            
            \item The intermediate models from all satellites are uploaded to the DPC, where they are aggregated into the updated global model $\mathbf{z}(t+1)$ through a second OTA-enabled hop.
        \end{enumerate}
        
Let $\mathcal{K}(t)\subseteq\mathcal{K}$ denote the set of participating devices at round $t$. 
Each participating device $k\in\mathcal{K}(t)$ performs $E$ local stochastic gradient descent epochs on its current local dataset $\mathcal{D}_k(t)$. 
Upon reception of the global model, device $k$ initializes
$\mathbf{w}_{k,0}^{(1)}(t) = \mathbf{z}(t).$
For local epoch $e=1,\dots,E$, device $k$ iterates over mini-batches from its local dataset $\mathcal{D}_k(t)$. 
Let $b=1,\dots,M_k(t)$ index the mini-batches in epoch $e$, where $M_k(t)$ denotes the number of mini-batches per epoch determined by $M_k(t)=\left\lceil |\mathcal{D}_k(t)|/B \right\rceil$. 
Let $\mathcal{B}_{k,b}^{(e)}(t)\subseteq\mathcal{D}_k(t)$ denote the $b$-th mini-batch of size $B$. 
The stochastic gradient at iteration $(e,b)$ is
\begin{equation}
\mathbf{g}_{k,b}^{(e)}(t)
=\nabla_{\mathbf{w}}
\left(
\frac{1}{B}\sum_{(\mathbf{x},y)\in \mathcal{B}_{k,b}^{(e)}(t)}
\ell\!\left(\mathbf{w}_{k,b-1}^{(e)}(t);\mathbf{x},y\right)
\right),
\end{equation}
where $\ell(\cdot)$ denotes the per-sample loss. The local model is then updated by
\begin{equation}
\mathbf{w}_{k,b}^{(e)}(t)
=\mathbf{w}_{k,b-1}^{(e)}(t)-\eta_l\,\mathbf{g}_{k,b}^{(e)}(t),
\end{equation}
where $\mathbf{w}_{k,0}^{(e)}(t)\triangleq \mathbf{w}_{k,M_k(t)}^{(e-1)}(t)$ for $e\ge 2$.
After $E$ local epochs, the updated local model is
\begin{equation}
\mathbf{z}_k(t+1)\triangleq \mathbf{w}_{k,M_k(t)}^{(E)}(t),
\end{equation}
which in practical deployments may be obtained at varying times across devices due to the heterogeneous computing capacities and varying available data among devices. To tackle this asynchronous issue and strictly guarantee the physical-layer synchronization required by the subsequent over-the-air aggregation, the system enforces a strict data buffer limit. Specifically, each device is configured with a software-defined data queue of a maximum capacity, keeping only a restricted number of the most recently collected samples to actively preserve data freshness. By bounding the maximum size of the local dataset, the computational workload per round is tightly capped. Consequently, all heterogeneous devices are forced to finish their local epochs within a short and uniform time window. Once finished, the devices cache their updated models and wait for a synchronization beacon broadcast by the satellite to concurrently transmit their signals in the uplink phase. This mechanism effectively decouples the asynchronous local computing from the synchronous OTA transmission, ensuring the aggregated signal is successfully received during the subsequent OTA uplink aggregation phase of round $t$. This abstract local update formulation applies to both the MLP and CNN models considered in Section \ref{SEC:IIII} and is agnostic to the specific network architecture.
Each device computes its local update based on a local objective. Let $\mathcal{D}_k(t)$ denote the local dataset of device $k$ at round $t$, and $|\mathcal{D}_k(t)|$ its cardinality. 
        We define the local loss function of device $k$ at round $t$ as a function of an arbitrary model parameter vector $\mathbf{w}$
        \begin{equation}
\mathcal{L}_k(\mathbf{w};t) = \frac{1}{|\mathcal{D}_k(t)|} \sum_{(\mathbf{x}, y) \in \mathcal{D}_k(t)} \ell(\mathbf{w}; \mathbf{x}, y).
\end{equation}
        Accordingly, the global objective of the FL process is to minimize a weighted sum of local losses across all participating devices, which can be defined as
\begin{equation}\label{eq:global_obj}
\underset{\mathbf{z}(t)\in\mathbb{R}^{d}}{\mathrm{minimize}}\quad
\mathcal{L}\big(\mathbf{z}(t)\big)
= \frac{1}{\psi(t)}\sum_{k\in\mathcal{K}(t)} \phi_k(t)\, \mathcal{L}_k\big(\mathbf{z}(t);t\big),
\end{equation}
        where $d$ is the dimension of the model parameter vector $\mathbf{z}(t)$, and
        $\phi_k(t)=\big|\mathcal{D}_k(t)\big|,
        \psi(t)=\sum_{i\in\mathcal{K}(t)}\big|\mathcal{D}_i(t)\big|$
        are the transmitter preprocessing scalar and the receiver postprocessing scalar, respectively. 
 We define the beam activation pattern of each satellite $n \in \mathcal{N}$ at round $t$ by a binary vector
\begin{equation}
\mathbf{X}_{n}(t)=\big[x_{n,1}(t),\cdots,x_{n,c}(t),\cdots,x_{n,C}(t)\big],
\end{equation}   
        where $x_{n,c}(t) \in \{0, 1\}$ indicates whether the $c$-th covered cell $s_{n,c}(t)$ is selected by satellite $n$ at round $t$. 
        The beam activation constraint requires that the number of active beams does not exceed the maximum, i.e., $\sum_{c=1}^{C} x_{n,c}(t) \leq V$, and the selected cells satisfy $\mathcal{A}_n(t) = \{s_{n,c}(t)\,|\,x_{n,c}(t)=1\}$. 
        With a slight abuse of notation, we let $c$ index the $c$-th covered cell $s_{n,c}(t)$ of satellite $n$, and write $\mathcal{K}_c \equiv \mathcal{K}_{s_{n,c}(t)}$ for its device set.

        \subsection{Channel Model}

Although each satellite transmits a single beam to a selected cell in the downlink, in the uplink direction, device signals can be received by all beams within the satellite's active beam set $\mathcal{V}_n(t)$ due to the broadcast nature of wireless propagation and the digital beamforming capability at the payload. 
We focus on the dominant line-of-sight (LoS) component and neglect shadowing and multipath effects, which is reasonable for LEO satellite links with clear sky conditions. 
We further assume that devices have compensated for Doppler shift caused by satellite motion through standard frequency-tracking techniques, and that rain attenuation and atmospheric absorption are negligible under the considered scenario.
        
        Let $G_k$ denote the transmit antenna gain of device $k$, and $G_v$ denote the receive antenna gain of satellite beam $v \in \mathcal{V}_n(t)$. The channel gain between device $k$ and beam $v$ at time $t$ is given by\cite{channel}
        \begin{equation}
        {{h}_{k,v}}\left( t \right)=\frac{\sqrt{{{G}_{k}}{{G}_{v}}}}{4\pi {{d}_{k,v}}\left( t \right)/ \lambda},
        \label{eq:chanelgain}
        \end{equation}
        where $d_{k,v}(t)$ is the distance between device $k$ and beam $v$, and $\lambda$ is the carrier wavelength. The satellite receive antenna pattern follows the 3GPP TR 38.811 directional model\cite{3gppTR38811}
        \begin{equation}
        G_v(\theta) =
        \begin{cases}
        G_m, & \theta = 0^\circ, \\
        G_m \left[ \frac{4 J_1(2\pi a \sin \theta / \lambda)}{2\pi a \sin \theta / \lambda} \right]^2, & 0^\circ < \theta \leq 90^\circ,
        \end{cases}
        \end{equation}
        where $\theta$ is the off-axis angle, $a$ is the aperture radius of the antenna, $J_1(\cdot)$ is the first-order Bessel function of the first kind, and $G_m$ is the maximum antenna gain, given by 
\begin{equation}
    {{G}_{m}}=10{{\log }_{10}}\left( \frac{\pi {{{{D}'}}^{2}}}{{{\lambda }^{2}}} \right),
\end{equation}
where $D' = 2a$ denotes the diameter of the antenna aperture. For the DPC and devices, we adopt a simplified directional antenna model
\begin{equation}
G_x(\theta) =
\begin{cases}
G_x^{\text{max}}, & \theta \leq \theta_{\text{main}}, \\
G_x^{\text{max}} - \Delta G(\theta), & \theta > \theta_{\text{main}},
\end{cases}
\end{equation}
where $x \in \{k, g\}$ denotes either a device or the DPC, and $\theta$ denotes the angle between the antenna boresight and the direction to the satellite beam, and the attenuation function $\Delta G(\theta)$ is given by
\begin{equation}
\Delta G(\theta) = k_a \left( \frac{\theta}{\theta_a} \right)^{n_a},
\end{equation}
where $k_a$ controls the initial attenuation and $n_a$ governs the decay sharpness. Thus, $\Delta G(\theta)$ captures the roll-off of side-lobe radiation. 
We assume the satellite transmit and receive antenna gains are identical, and that the satellite aligns its antenna boresight toward the DPC during the upload of intermediate models. 
The channel gain between satellite $n \in \mathcal{N}$ and the DPC at round $t$ is given by
        \begin{equation}
        h_{n,g}(t) = \frac{\sqrt{G_m G_g}}{ 4\pi d_{n,g}(t) / \lambda },
        \end{equation}
        where $d_{n,g}(t)$ is the distance between satellite $n$ and the DPC.
        \subsection{OTA Computation} 
        \subsubsection{Device-to-LEO Uplink}
        Let $\mathcal{K}_n(t) \subseteq \mathcal{K}(t)$ denote the subset of participating devices served by satellite $n$ at round $t$. Specifically, $\mathcal{K}_n(t)$ comprises the active devices located within the cells covered by the activated beams of satellite $n$.
        Define the normalized signal vector $\mathbf{s}_k(t) \in \mathbb{R}^d$ from device $k$ such that $\mathbb{E}\{\mathbf{s}_k(t) \mathbf{s}_k^H(t)\} = \mathbf{I}$, where each entry $s_k^{(i)}(t)$ corresponds to the $i$th dimension of the (scaled) local model update $\mathbf{z}_k(t)$. 
        In each dimension $i \in \{1, \dots, d\}$, the received signal at satellite $n$ considering channel gains and noise is
        \begin{equation}
        \hat{r}^{(i)}_n(t) = \sum_{k \in \mathcal{K}_n(t)} \sum_{v \in \mathcal{V}_n(t)} h_{k,v}(t) b_k(t) s_k^{(i)}(t) + \varepsilon_n^{(i)}(t),
        \end{equation}
        where $b_k(t)$ is the transmit power coefficient of device $k$, $\varepsilon_n^{(i)}(t) \sim \mathcal{CN}(0, \sigma_n^2)$ is additive white Gaussian noise (AWGN). The noise variance $\sigma_n^2$ models the receiver thermal noise at each satellite, and the per-device power constraint in \eqref{eq:power-constraint} ensures that each device respects its maximum transmit budget, i.e.,
        \begin{equation}
        \mathbb{E}\left( {{\left| {{b}_{k}}\left( t \right){{s}_{k}}\left( t \right) \right|}^{2}} \right)={{\left| {{b}_{k}}\left( t \right) \right|}^{2}}\le {{P}_{k}}.
        \label{eq:power-constraint}
        \end{equation}

        \subsubsection{LEO-to-DPC Uplink}
        
        Under ideal conditions, the final global model is obtained by aggregating the local models from all participating devices, which can be written as
        \begin{equation}
        \mathbf{z}\left( t \right)=\frac{1}{\psi \left( t \right)}\sum\limits_{k\in \mathcal{K}\left( t \right)}{{{\phi }_{k}}\left( t \right)\mathbf{z}_{k}^{{}}\left( t \right)}.
        \end{equation}
To preserve the structure of this weighted sum over a two-hop OTA link, we set the transmit vector of satellite $n$ as the unit-variance normalization of its received OTA sum
        \begin{equation}
        \mathbf{s}_n(t) \triangleq 
        \frac{\hat{\mathbf{r}}_n(t)}
        {\sqrt{\mathbb{E}\{\|\hat{\mathbf{r}}_n(t)\|^2\}/d}}, 
        \mathbb{E}\{\mathbf{s}_n(t)\mathbf{s}_n^H(t)\}=\mathbf{I},
        \end{equation}
        so that the transmit power control is captured by the scalar $b_n(t)$. 
        For brevity, we omit the timeslot index $i$ in subsequent expressions. 
        The received signal at the DPC is
        \begin{equation}
        \hat{r}_g(t) = \sum_{n \in \mathcal{N}} h_{n,g}(t) b_n(t) s_n(t) + \varepsilon_g(t),
        \end{equation}
        where $b_n(t)$ is the transmit power coefficient of satellite $n$, $\varepsilon_g(t) \sim \mathcal{CN}(0, \sigma_g^2)$ is AWGN, and the power constraint is
        \begin{equation}
        \mathbb{E}\left( {{\left| {{b}_{n}}\left( t \right)s_{n}^{{}}\left( t \right) \right|}^{2}} \right)={{\left| {{b}_{n}}\left( t \right) \right|}^{2}}\le {{P}_{n}}.
        \end{equation}

After normalization at the DPC, an estimate $\hat{\mathbf{z}}(t)$ of the desired global model $\mathbf{z}(t)$ is recovered from $\hat{r}_g(t)$. 
The dual-layer OTA design thus realizes an approximate analog implementation of the ideal weighted averaging rule across devices and satellites. The distortion between $\mathbf{z}(t)$ and $\hat{\mathbf{z}}(t)$ is evaluated by the MSE, which is given by
        \begin{equation}
        \label{eq:MSE}
        \begin{aligned}
        & \mathrm{MSE}\bigl(\hat{\mathbf z}(t),\mathbf z(t)\bigr)=\mathbb{E}\left(\left\lVert \hat{\mathbf z}(t)-\mathbf z(t) \right\rVert^{2}\right)\!=\!\sum\limits_{n\in \mathcal{N}}{\sum\limits_{c\in {{\mathcal{S}}_{n}}\left( t \right)}\!{x_{n,c}^{{}}\left( t \right)\cdot }} \\ 
         & \sum\limits_{{{k}_{{}}}\in {{\mathcal{K}}_{c}}}{{{\left| \frac{{{h}_{n,g}}\left( t \right){{b}_{n}}\left( t \right)\sum\limits_{v\in {{\mathcal{V}}_{n}}\left( t \right)}{{{h}_{k,v}}\left( t \right){{b}_{k}}\left( t \right)}}{\sqrt{\mathbb{E}\left( {{\left\| {{\widehat{\mathbf{r}}}_{g}}(t) \right\|}^{2}} \right)\mathbb{E}\left( {{\left\| {{\widehat{\mathbf{r}}}_{n}}(t) \right\|}^{2}} \right)}/d}-\frac{{{\phi }_{k}}\left( t \right)}{\psi \left( t \right)} \right|}^{2}}} \\ 
         & +\sum\limits_{n\in \mathcal{N}}{\frac{h_{n,g}^{2}\left( t \right)b_{n}^{2}\left( t \right)\sigma _{n}^{2}}{\mathbb{E}\left( {{\left\| {{\widehat{\mathbf{r}}}_{g}}(t) \right\|}^{2}} \right)\mathbb{E}\left( {{\left\| {{\widehat{\mathbf{r}}}_{n}}(t) \right\|}^{2}} \right)/{{d}^{2}}}}+\frac{\sigma _{g}^{2}}{\mathbb{E}\left( {{\left\| {{\widehat{\mathbf{r}}}_{g}}(t) \right\|}^{2}} \right)} .
        \end{aligned}
        \end{equation}
The first term represents the bias error caused by the mismatch between the effective analog combining weights induced by channels and power coefficients $\{h_{k,v}(t),b_k(t),h_{n,g}(t),b_n(t)\}$ and the desired digital aggregation weights $\{\phi_k(t)/\psi(t)\}$. 
This term captures the residual distortion that remains even in the absence of noise when the analog scaling cannot perfectly realize the target weighted average. 
The second term corresponds to the uplink noise amplification from devices to satellites: the receiver noise $\sigma_n^2$ at each satellite is first added in $\hat{\mathbf{r}}_n(t)$ and is then re-scaled by the subsequent OTA forwarding, so that its contribution to the global model depends on both $h_{n,g}(t)$ and $b_n(t)$ as well as the normalization factors. 
The third term accounts for the ground-side noise at the DPC, which directly perturbs the aggregated signal $\hat{r}_g(t)$ regardless of the uplink configuration and sets a fundamental noise floor. 
Together, these three components illustrate how the choice of beam-hopping pattern and transmit powers jointly influences both the aggregation bias and the noise-induced distortion, thereby linking physical-layer control actions to learning accuracy.

\subsection{Optimization Problem Formulation}
Let $|\bar{\mathcal{D}}_{k}(t)|$ denote the amount of new local training data collected by device $k$ in round $t$. To preserve data freshness and enforce a strictly bounded computational workload for physical layer synchronization, each device maintains a software-defined data buffer with a maximum capacity of $D_{\max}$. Taking into account the cached data from previous rounds when a device was not selected, the available data of device $k$ at round $t$ is mathematically defined as
\begin{equation}
|{\mathcal{D}}_{k}(t)| = \min \Big( D_{\max}, |{\bar{\mathcal{D}}}_{k}(t)| + (1 - a_{k}(t-1)) \lfloor \eta |{\mathcal{D}}_{k}(t-1)| \rfloor \Big),
\label{eq:cached-data-discount}
\end{equation}
where $a_k(t)\in\{0,1\}$ is the selection indicator (with $a_k(t)=1$ if device $k$ participates in round $t$ and $a_k(t)=0$ otherwise), which is mapped from the cell activation variables as $a_k(t) = \sum_{n=1}^{N} x_{n,c}(t)$ for $k \in \mathcal{K}_c$. Furthermore, $\eta \in (0,1]$ is a constant discount factor characterizing the loss of freshness of previously untrained data. 
The floor operator ensures that the buffered data amount remains integer-valued at the sample level, and the $\min$ operator explicitly enforces the buffer capacity constraint to drop stale data when the queue overflows.
This data arrival and freshness-aware buffering mechanism distinguishes the considered system from conventional offline FL where local datasets are static, and links the device scheduling decisions with how much previously collected (and gradually aging) data can be effectively utilized.
The total amount of training data actually exploited in round $t$ is
        \begin{equation}
        |\mathcal{D}\left( t \right)|=\sum\limits_{n=1}^{N}{\sum\limits_{c=1}^{C}{x_{n,c}^{{}}\left( t \right)\sum\limits_{k=1}^{\left| {{\mathcal{K}}_{c}} \right|}{\left| {{\mathcal{D}}_{k}}\left( t \right) \right|}}}.
        \label{eq:total-data}
        \end{equation}
It is worth noting that we assume the local data collected by ground devices are independent and identically distributed (IID) across the network. Under this assumption, the local datasets inherently contain a diversity of classes and represent the true global distribution. Consequently, maximizing the total amount of participating data mathematically reduces the variance of the stochastic gradients and directly contributes to minimizing the global loss. 
Due to beam, power, and MSE constraints, it is infeasible to aggregate all device updates in a single round. In addition, closely spaced beams from different satellites may result in severe inter-satellite interference. 
Hence, adaptive BH design is required to allocate satellite beams dynamically in a way that minimizes interference and optimizes data utilization. Our objective is to maximize the long-term total amount of training data exploited by the OTA-enabled online FL process, which is formalized as follows:
\begin{subequations}\label{eq:P0}
  \begin{equation}\label{eq:P0-obj}
    \underset{{}}{\mathop{\underset{\{{{\mathbf{X}}_{n}}(t),{{b}_{k}}(t),{{b}_{n}}(t)\}}{\mathop{\text{maximize}}}\,}}\,\ \ \frac{1}{T}\sum\limits_{t=1}^{T}{\left| \mathcal{D}\left( t \right) \right|},
    \tag{\theparentequation} 
  \end{equation}
  \setcounter{equation}{0}
  \begin{align}
 & \text{s}\text{.t}\text{.}\ \ x_{n,c}^{{}}\left( t \right)\in \left\{ 0,1 \right\},\forall n\in \mathcal{N},c\in {{\mathcal{S}}_{n}}\left( t \right),t\in \mathcal{T} , \label{22a}\\ 
 & \quad \ \ \sum\limits_{n=1}^{N}{x_{n,c}^{{}}\left( t \right)}\le 1,\forall c\in \mathcal{S},t, \label{22b}\\ 
 & \quad \ \ \sum\limits_{c=1}^{C}{x_{n,c}^{{}}\left( t \right)}\le V,\forall n\in \mathcal{N},t, \label{22c}\\ 
 & \quad \ \ x_{n,c}^{{}}\left( t \right)=0,\forall n\in \mathcal{N},c\notin {{\mathcal{S}}_{n}}\left( t \right),t, \label{22d}\\ 
 & \quad \ \ {{\omega }_{i,j}}\ge \bar{\omega },\text{if}\ x_{n,i}^{{}}\left( t \right)x_{n',j}^{{}}\left( t \right)=1,\forall n,n'\in \mathcal{N},i,j,t, \label{22e}\\ 
 & \quad \ \ {{\left| {{b}_{k}}\left( t \right) \right|}^{2}}\le {{P}_{k}}\sum\limits_{n=1}^{N}{x_{n,c}^{{}}\left( t \right)},\forall c\in \mathcal{S},k\in {{\mathcal{K}}_{c}},t, \label{22f}\\
 & \quad \ \ {{\left| {{b}_{n}}\left( t \right) \right|}^{2}}\le {{P}_{n}},\forall n\in \mathcal{N},t, \label{22g}\\ 
 & \quad \ \ \text{MSE}\left( \mathbf{\hat{z}}\left( t \right),\mathbf{z}\left( t \right) \right)\le \rho ,\forall t, \label{22h}
  \end{align}
\end{subequations}
where $\mathbf{X}_n(t)=[x_{n,1}(t),\ldots,x_{n,C}(t)]$ is the binary beam activation vector of satellite $n$, and $b_k(t)$ and $b_n(t)$ denote the transmit power coefficients of device $k$ and satellite $n$, respectively. 
Specifically, constraints \eqref{22a} and \eqref{22b} define the binary beam activation variables and ensure that each cell is served by at most one satellite to avoid simultaneous illumination conflicts.
Constraints \eqref{22c} and \eqref{22d} capture the satellite hardware and deployment limitations, where the former imposes the per-satellite beam budget $V$ and the latter restricts beam activation to the instantaneous visible footprint $\mathcal{S}_n(t)$.
To mitigate inter-satellite interference between closely spaced beams, constraint \eqref{22e} enforces a minimum distance $\bar{\omega}$ between active beams of different satellites.
Regarding power control, constraint \eqref{22f} specifies the maximum transmit power limit for ground devices and explicitly couples the device transmit power with the beam activation status, ensuring that $b_k(t)$ is strictly zero if its corresponding cell $c$ is not selected by any satellite. Constraint \eqref{22g} specifies the maximum transmit power limit for LEO satellites.
Finally, constraint \eqref{22h} sets an upper bound $\rho$ on the global aggregation MSE, which guarantees that the distortion introduced by the dual-layer OTA aggregation remains within a tolerable level for maintaining FL convergence.

The optimization problem \eqref{eq:P0} aims to maximize the long-term data utilization under strict physical and learning constraints.
It is formulated as a MINLP problem, characterized by the intricate coupling between the binary BH decisions and the continuous power control variables.
The complexity arises not only from the exponential growth of the discrete search space but also from the highly non-convex nature of the objective function and the global MSE constraint in \eqref{22h}, rendering the problem NP-hard.
Crucially, the time-varying satellite topology and dynamic data generation transform this optimization into a sequential decision-making problem, for which conventional static algorithms are computationally intractable.
In the next section, we reformulate the problem as an MDP and develop a PPO-based DRL approach that learns near-optimal BH and power control policies through interaction with the FL environment.

\section{Proximal Policy Optimization-Based DRL Framework}
\label{SEC:III}
\subsection{MDP Formulation}
\label{sec:mdp}
To tackle the MINLP problem in \eqref{eq:P0}, we reformulate the optimization as an MDP, which provides a suitable framework for sequential decision-making under system dynamics and long-term performance metrics \cite{BH2}. 
Specifically, the Markov property strictly holds in our formulation. The next state $o_{t+1}$ comprises the updated data queues and the new channel realizations. Its transition depends entirely on the current state $o_t$ and the scheduling action $a_t$ taken in round $t$, and is independent of the past historical states. The global model $\mathbf{z}(t)$ is accordingly excluded from $o_t$, because under the unit-variance normalization of Section~II-C the reward in \eqref{eq:reward_def} and the transitions of the data queues and channel gains depend only on the data amounts and channel gains, so that $\mathbf{z}(t)$ is driven by the scheduling decisions without feeding back into them. This intrinsic property guarantees that the DRL agent can learn the optimal policy based solely on current observations. 
At each round $t\in\mathcal{T}$, the scheduler observes a state $o_t$, selects an action $a_t$, and then receives a scalar reward $r_t$ determined by the resulting data utilization and aggregation distortion.


\textit{1) State:} 
The system state at round $t$ is defined as
\begin{equation}
\begin{aligned}
  & {{o}_{t}}=\left[ {{\mathbf{D}}_{t}};\text{vec}\left( {{\mathbf{H}}_{t}} \right);\mathbf{h}_{t}^{n,g} \right], \\ 
 & \left\{ \begin{aligned}
  & {{\mathbf{D}}_{t}}={{\left[ \begin{matrix}
   \left| \mathcal{D}_{t}^{1} \right| & \left| \mathcal{D}_{t}^{2} \right| & \cdots  & \left| \mathcal{D}_{t}^{K} \right|  \\
\end{matrix} \right]}^{\text{T}}}, \\ 
  & {{\mathbf{H}}_{t}}=\left[ \begin{matrix}
   \mathbf{h}_{t}^{1,1} & \mathbf{h}_{t}^{1,2} & \cdots  & \mathbf{h}_{t}^{1,C}  \\
   \mathbf{h}_{t}^{2,1} & \mathbf{h}_{t}^{2,2} & \cdots  & \mathbf{h}_{t}^{2,C}  \\
   \cdots  & \cdots  & \cdots  & \cdots   \\
   \mathbf{h}_{t}^{N,1} & \mathbf{h}_{t}^{N,2} & \cdots  & \mathbf{h}_{t}^{N,C}  \\
\end{matrix} \right], \\ 
  & \mathbf{h}_{t}^{n,g}={{\left[ \begin{matrix}
   h_{t}^{1,g} & h_{t}^{2,g} & \cdots  & h_{t}^{N,g}  \\
\end{matrix} \right]}^{\text{T}}},
\end{aligned} \right.
\end{aligned}
\end{equation}
where $|\mathcal{D}_{t}^{k}|$ is the available data amount of device $k$ obtained from \eqref{eq:cached-data-discount}, 
$\mathbf{h}_{t}^{n,c}={{\left[ \begin{matrix}
   h_{t}^{n1,c} & h_{t}^{n2,c} & \cdots  & h_{t}^{nC,c}  \\
\end{matrix} \right]}^{\text{T}}}$
represents the channel gains from the $C$ beams of satellite $n$ to the corresponding covered cells in round $t$, which can be calculated according to \eqref{eq:chanelgain}, and $\mathbf{h}_{t}^{n,g}$ represents the channel gains between satellites and the DPC.
The operator $\mathrm{vec}(\cdot)$ vectorizes ${\mathbf{H}}_{t}$ into a single column vector, thereby ensuring that $o_t$ has a fixed dimension for DRL implementation. 
This state definition jointly encodes the distribution of available data across devices, the instantaneous satellite–device channel conditions, and the satellite–DPC links, thereby capturing the key factors that influence both data utilization and OTA aggregation quality.

\textit{2) Action:} 
The action at round $t$ is
\begin{equation}
{{a}_{t}}=\{{{\{\mathbf{X}_{t}^{n}\}}_{n\in \mathcal{N}}},\ {{\{b_{t}^{k}\}}_{k\in \mathcal{K}}},\ {{\{b_{t}^{n}\}}_{n\in \mathcal{N}}}\},
\end{equation}
where $\mathbf{X}_{t}^{n}$ is the binary beam activation vector of satellite $n$, $b_{t}^{k}$ is the transmit power coefficient of device $k$, and $b_{t}^{n}$ is the transmit power coefficient of satellite $n$. 
In the proposed PPO framework, the actor network outputs a parameterization of the joint distribution over these continuous and discrete control variables, from which actions are sampled during training.

\textit{3) Reward:} 
We design the per-round reward to encourage utilizing more on-time data while penalizing aggregation distortion beyond the threshold. Specifically,
\begin{equation}
\label{eq:reward_def}
{{r}_{t}}=\frac{\left| {{\mathcal{D}}_{t}} \right|}{{{\left| {{\mathcal{D}}_{t}} \right|}_{\max }}}-\text{clip}\left( \mu \left( {{\text{MSE}}_{t}}-\rho  \right),0,1 \right),
\end{equation}
where $|{\mathcal{D}}_{t}|$ is the total amount of training data utilized in round $t$, ${{\left| {{\mathcal{D}}_{t}} \right|}_{\max }}$ is a normalization constant denoting the maximum attainable data amount in a single round, $\rho$ is the MSE threshold, and $\mu>0$ is the penalty weight. 
Here, $\text{MSE}_t \triangleq \text{MSE}(\hat{\mathbf{z}}(t), \mathbf{z}(t))$ denotes the instantaneous aggregation distortion derived in \eqref{eq:MSE}.
The function $\mathrm{clip}(x,0,1)\triangleq \min\{\max\{x,0\},1\}$ bounds the penalty so that no penalty is incurred when $\text{MSE}_t \le \rho$, while a capped penalty is applied otherwise. 

The first term in \eqref{eq:reward_def} structurally aligns the reward with the long-term objective defined in \eqref{eq:P0}, while the second term enforces the non-convex global MSE constraint in a soft manner by discouraging actions that push the aggregation error above $\rho$. 
Crucially, this reward design ensures strict alignment with the original MINLP objective. By maximizing the cumulative discounted reward $\sum_{t} \gamma^t r_t$, the DRL agent inherently maximizes the long-term utilized data amount, which is the exact objective of the original problem, while dynamically learning to satisfy the physical-layer aggregation constraints. Therefore, finding the optimal policy for this MDP directly yields a near-optimal solution to the intractable MINLP problem.
By normalizing both the data-utilization term and the penalty, the per-step reward remains bounded, which is beneficial for stabilizing PPO training and improving sample efficiency.

Under this MDP formulation, the goal of the DRL agent is to learn a stationary policy that maximizes the expected discounted cumulative reward, thereby implicitly approximating the original long-term data-utilization objective with MSE guarantees over time.

\subsection{PPO for OTA-Enabled Online FL}

\begin{figure}[t]
\centering
\includegraphics[width=0.5\textwidth]{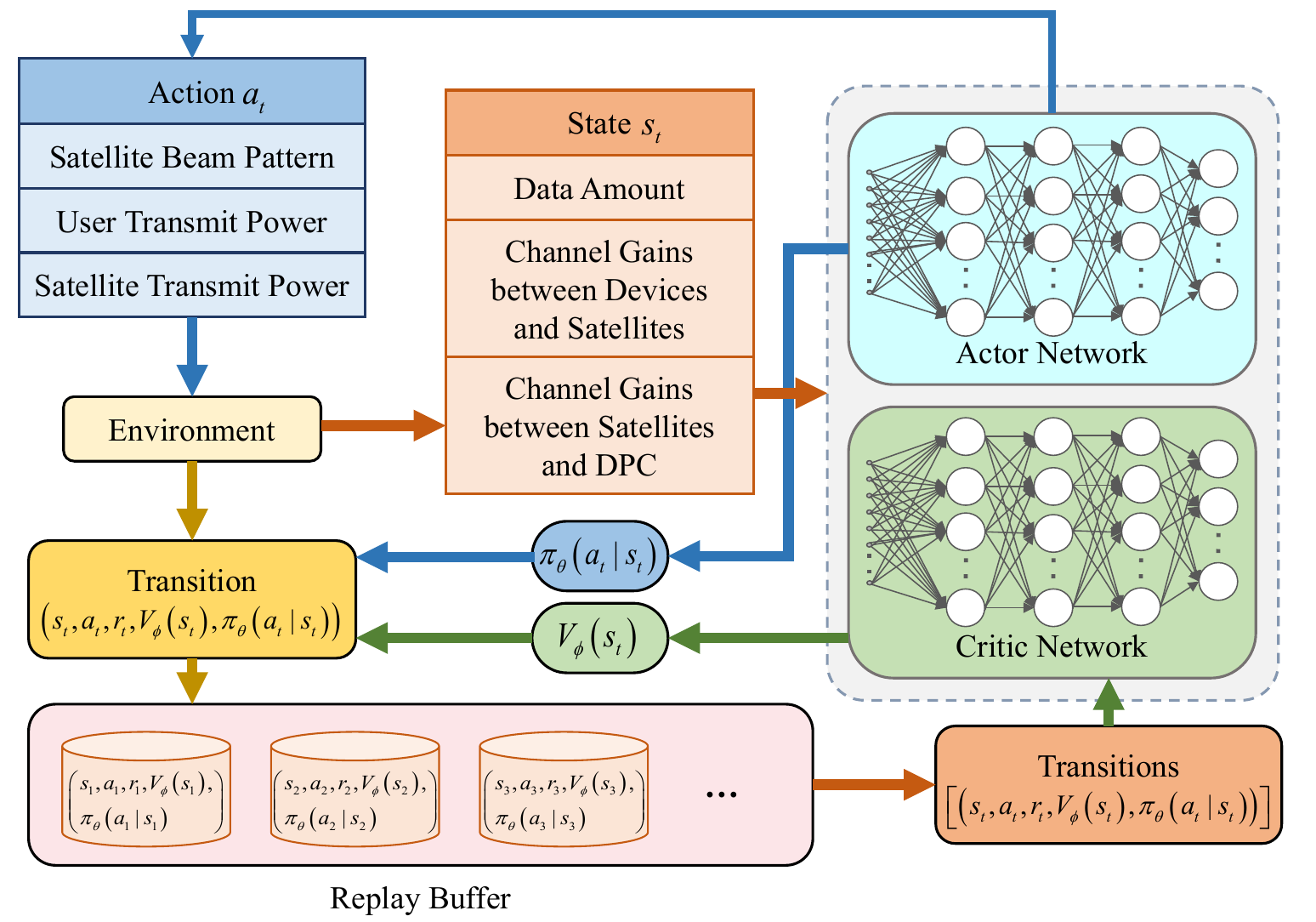}
\caption{PPO-based DRL approach for OTA-enabled online FL.}
\label{fig:ppo-architecture}
\end{figure}

To solve the MDP formulated in Section~\ref{sec:mdp}, we adopt a model-free policy gradient method based on PPO~\cite{ppo}. 
PPO learns a parameterized stochastic policy that maximizes the long-term return while maintaining stable updates through a clipped surrogate objective, which is well suited to our high-dimensional action space composed of beam activation and power control variables.

The overall PPO-based scheduling architecture is illustrated in Fig.~\ref{fig:ppo-architecture}. 
At each global round $t$, the environment produces the state $o_t$, which aggregates the data-amount and channel-gain information defined in Section~\ref{sec:mdp}. 
The actor network takes $o_t$ as input and outputs a stochastic policy $\pi_{\theta}(a_t|o_t)$ over the action $a_t$, whose components are the beam-hopping decision and the transmit-power coefficients of devices and satellites. 
In parallel, the critic network outputs the value estimate $V_{\psi}(o_t)$. 
A raw action $\tilde a_t$ is sampled from $\pi_{\theta}(\cdot|o_t)$ and then mapped to a feasible action $a_t=\mathcal{M}(\tilde a_t)$ that satisfies the system constraints (e.g., beam-budget and conflict constraints) before being applied to the OTA-enabled online FL environment. 
The environment then transitions to the next state $o_{t+1}$ and yields the per-round reward $r_t$. 
The resulting transitions are stored in a replay buffer and used to update the actor and critic; the buffer is cleared after each update.

For notational clarity, we denote a trajectory collected under policy $\pi_{\theta}$ as 
$\tau = \{(o_t,a_t,r_t,o_{t+1})\}_{t=0}^{T-1}$. 
The discounted return starting from round $t$ is
\begin{equation}
G_t = \sum_{l=0}^{T-1-t} \gamma^{l} r_{t+l},
\label{eq:return}
\end{equation}
where $0<\gamma<1$ is the discount factor and $r_t$ is the per-round reward defined in~\eqref{eq:reward_def}. The state-value function under policy $\pi_{\theta}$ is
\begin{equation}
V^{\pi_{\theta}}(o_t) = \mathbb{E}_{\pi_{\theta}}\!\left[ G_t \,\big|\, o_t \right],
\label{eq:value}
\end{equation}
which is approximated by the critic network $V_{\psi}(o_t)$ with parameters $\psi$. 
Since the action includes both discrete beam-hopping decisions and continuous power coefficients, we adopt a hybrid parameterization of $\pi_{\theta}(a_t|o_t)$. 
Specifically, the actor outputs (i) real-valued beam scores that are subsequently mapped to a binary beam activation vector by a feasibility mapping, and (ii) continuous power coefficients modeled by a diagonal Gaussian distribution whose mean and variance are produced from $o_t$. 

The mapping $a_t=\mathcal{M}(\tilde a_t)$ is deterministic and enforces the beam and conflict constraints of \eqref{eq:P0} in sequence. The beam scores are first binarized by a fixed threshold. Because adjacent footprints overlap, a cell may be claimed by several satellites, in which case only the satellite attaining the highest beam score retains it and the remaining candidates are switched off, which enforces \eqref{22b}. If a satellite still activates more than $V$ beams, only the $V$ highest-scoring ones are kept, which enforces \eqref{22c}. The power variables are linearly rescaled onto the admissible intervals so that \eqref{22f} and \eqref{22g} hold, and $b_k(t)$ is set to zero for every device whose cell is not activated. Since $\mathcal{M}$ is applied on the environment side, the log-probability stored for the PPO update is that of the raw action $\tilde a_t$, and the ratio in \eqref{eq:ratio} is formed by evaluating $\pi_{\theta}$ and $\pi_{\theta_{\mathrm{old}}}$ at the same $\tilde a_t$.
To encourage sufficient exploration of diverse BH and power-control patterns, we use the policy entropy at round $t$
\begin{equation}
\mathcal{H}_t \triangleq \mathcal{H}\!\big(\pi_{\theta}(\cdot|o_t)\big)
= \mathbb{E}_{a_t \sim \pi_{\theta}}\!\big[-\log \pi_{\theta}(a_t|o_t)\big],
\label{eq:entropy}
\end{equation}
as an additional regularization term in the policy objective. 
Directly estimating the policy gradient from returns $G_t$ usually suffers from high variance. 
Therefore, we adopt generalized advantage estimation (GAE), which constructs a low-variance advantage estimator from temporal-difference (TD) residuals. Let
\begin{equation}
\delta_t = r_t + \gamma V_{\psi}(o_{t+1}) - V_{\psi}(o_t)
\label{eq:td_residual}
\end{equation}
denote the one-step TD error at round $t$. The GAE-based advantage estimate is given by
\begin{equation}
\hat{A}_t = \sum_{l=0}^{T-1-t} (\gamma \lambda_{\text{GAE}})^{l} \,\delta_{t+l},
\label{eq:gae}
\end{equation}
where $0 \le \lambda_{\text{GAE}} \le 1$ controls the bias-variance tradeoff. Here $\lambda_{\text{GAE}}$ always carries the subscript ``GAE'' and should not be confused with the carrier wavelength $\lambda$ used in the channel model of Section~II-A. 
Before optimization, $\{\hat{A}_t\}$ are normalized over a minibatch to improve numerical stability. 
PPO constrains policy updates by clipping the importance sampling ratio between the new and old policies. Given samples generated by the old policy $\pi_{\theta_{\mathrm{old}}}$, the ratio at round $t$ is
\begin{equation}
\zeta_t(\theta) = 
\frac{\pi_{\theta}(a_t|o_t)}{\pi_{\theta_{\mathrm{old}}}(a_t|o_t)}.
\label{eq:ratio}
\end{equation}

\begin{algorithm2e}[t]
\caption{PPO-based DRL approach for OTA-enabled online FL}
\label{alg:ppo}
\DontPrintSemicolon
\KwIn{Total episodes $N_{\text{ep}}$, horizon $T$, update interval $K_{\mathrm{PPO}}$, Initial environment $\mathcal{E}$, actor $\pi_{\theta}$, critic $V_{\psi}$, global FL model $\mathbf{z}(0)$.}
\KwOut{Optimized policy $\pi_{\theta^\star}$ yielding BH $\{\mathbf{X}_n(t)\}$ and power control $\{b_k(t),b_n(t)\}$.}
Initialize replay buffer $\mathcal{B}\leftarrow \varnothing$, and global step counter $m\leftarrow 0$.\;
\For{episode $=1,\dots,N_{\text{ep}}$}{
Initialize environment $\mathcal{E}$, reset global model $\mathbf{z}(0)$, and clear all device data buffers.\;
Observe initial state $o_0$.\; 
  \For{$t=0,\ldots ,T\!-\!1$}{
    Sample raw action $\tilde a_t=\{\{\tilde x_{n,c}(t)\},\{\tilde b_k(t)\},\{\tilde b_n(t)\}\}\sim\pi_{\theta}(\cdot|o_t)$.\;
    Map $\tilde a_t$ to a feasible action $a_t=\mathcal{M}(\tilde a_t)$ via score thresholding, conflict resolution by highest score, and top-$V$ beam selection.\;
    Run one FL round with $a_t$ in $\mathcal{E}$ to obtain updated $\mathbf{z}(t+1)$ and $o_{t+1}$.\;
    Compute reward $r_t=|\mathcal{D}\left( t \right)|/|\mathcal{D}|_{\max}-\text{clip}\left( \mu \left( \text{MSE}\left( t \right)-\rho  \right),0,1 \right)$.\;
    Store transition $(o_t,\tilde a_t,r_t,o_{t+1},\log\pi_{\theta_{\mathrm{old}}}(\tilde a_t|o_t),V_{\psi}(o_t))$ in $\mathcal{B}$.\;
    Synchronize global model $\mathbf{z}(t+1)$.\; 
    Set $o_t \leftarrow o_{t+1}$, $m \leftarrow m+1$.\;
  
  \If{m mod $K_{\mathrm{PPO}}$ $= 0$}{
  Compute advantages $\{\hat A\}$ with GAE~\eqref{eq:gae} and normalize.\; 
  Compute value targets $\{\hat V\}$ via~\eqref{eq:v_target}.\;
    Update actor by maximizing~\eqref{eq:ppo_obj} with entropy regularization using samples in $\mathcal{B}$ and apply gradient clipping.\;
    Update critic by minimizing~\eqref{eq:v_loss}.\;
    Set old policy parameters $\theta_{\text{old}} \leftarrow \theta$.\;
  }
  }
}
\end{algorithm2e}

The clipped surrogate objective with entropy regularization is
\begin{equation}
\begin{aligned}
J(\theta) 
= 
&\mathbb{E}_t \Big[
\min\big( \zeta_t(\theta)\hat{A}_t,\;
\mathrm{clip}\big(\zeta_t(\theta), 1-\epsilon, 1+\epsilon\big)\hat{A}_t \big)
\\[-1mm]
&
+ \, c_{\mathrm{e}}\,\mathcal{H}_t
\Big],
\end{aligned}
\label{eq:ppo_obj}
\end{equation}
where $\epsilon>0$ controls the update trust region, $c_{\mathrm{e}}\!>\!0$ is the entropy weight, and $\mathrm{clip}(\cdot)$ truncates the ratio into $[1-\epsilon,1+\epsilon]$. The clipping operation in~\eqref{eq:ppo_obj} prevents excessively large policy updates that could degrade performance, while the entropy term encourages exploratory beam and power configurations. For value function learning, we construct a target value $\hat{V}_t$ from the advantage estimator
\begin{equation}
\hat{V}_t = \hat{A}_t + V_{\psi}(o_t),
\label{eq:v_target}
\end{equation}
and minimize the mean-squared error between the critic output and target
\begin{equation}
L^{\text{v}}(\psi) = \mathbb{E}_t\Big[\big(V_{\psi}(o_t) - \hat{V}_t\big)^2\Big].
\label{eq:v_loss}
\end{equation}
In practice, the actor and critic are updated jointly by stochastic gradient descent on the composite loss
\begin{equation}
L(\theta,\psi) = -J(\theta) + c_{\mathrm{v}} L^{\text{v}}(\psi),
\label{eq:joint_loss}
\end{equation}
where $c_{\mathrm{v}}>0$ balances the policy and value learning terms. Gradient clipping is further applied to avoid exploding gradients in highly non-stationary environments.

The complete PPO-based DRL approach for OTA-enabled online FL is summarized in \textbf{Algorithm~\ref{alg:ppo}}.
Note that, at the start of every episode, the environment initialization empties all device data buffers together with resetting the global model, so that each episode begins from an independent initial condition and the initial state of one episode does not depend on the termination state of the previous one.

\section{Simulation Results}
\label{SEC:IIII}
\subsection{Experimental Setup and Benchmark Algorithms}
In this section, we verify the performance of the proposed PPO-based algorithm through numerical simulations, comparing it against several baseline algorithms. Specifically, we evaluate the proposed approach in terms of DRL training efficiency and stability, FL convergence performance, and long-term data utilization capability under varying system configurations.
The main system parameters are summarized in Table~\ref{tab:sys-params}~\cite{satfed2,BH2}, including orbital altitude, coverage radius, bandwidth, the device/satellite power budgets, and the reward function parameters such as the MSE penalty weight.

To evaluate the performance and robustness of the proposed PPO-based DRL approach under different workload complexities, we consider two representative FL tasks.
The first task is handwritten digit recognition on the MNIST dataset, which contains 60{,}000 training and 10{,}000 test images. 
Each device is assigned a unique, mutually disjoint shard of the training set. The shards are formed by letting the devices in turn draw equal-size subsets of sample indices uniformly at random without replacement from the pool of not-yet-assigned indices, which is statistically equivalent to shuffling the whole training set once and splitting it into consecutive blocks of equal size. Since no label information is used at any stage, the shards are neither sorted nor grouped by class, and the label distribution of each device is a multinomial draw with the global class probabilities, which is exactly the IID setting assumed in Section~II-C and reported in Table~III. The partition is fixed throughout training, and in each round a participating device draws the $|\mathcal{D}_k(t)|$ samples used for local training uniformly at random from its own shard. The same procedure is applied to CIFAR-10.
The global model is a multi-layer perceptron (MLP) with several fully connected layers and ReLU activations, trained for 60 global communication rounds with 2 local epochs.
The second task is image classification on the CIFAR-10 dataset, which consists of 50{,}000 training and 10{,}000 test color images. 
For this task, we adopt a convolutional neural network (CNN) with multiple convolutional and fully connected layers, extending the training process to 180 global rounds with 3 local epochs.
All FL-related hyperparameters and model specifications for both tasks are listed in Table~\ref{tab:fl-params}. 
These two tasks jointly allow us to assess not only the performance of the proposed algorithm on a standard benchmark, but also its robustness under more complex learning workloads.

The DRL agent that controls BH and transmit power is implemented using the PPO algorithm. 
The actor and critic networks both consist of fully connected layers with ReLU activations. 
The agent is trained over 1,500 episodes to ensure convergence.
We adopt task-specific PPO training hyperparameters summarized in Table~\ref{tab:drl-params}.
Key parameters include a discount factor of $\gamma=$ 0.95, a GAE parameter of $\lambda_{\text{GAE}}=$ 0.95, and a PPO clip ratio of $\epsilon=$ 0.2 to maintain training stability.
The actor network learning rate is set to $\alpha_{\mathrm{a}}=5\times 10^{-4}$ for MNIST and $1\times 10^{-3}$ for CIFAR-10, while the critic learning rate is fixed at $\alpha_{\mathrm{c}}=2\times 10^{-3}$.
The PPO update interval is set to $K_{\mathrm{PPO}}=256$ steps, which triggers a policy update once every $256$ collected transitions. This value balances the freshness of the on-policy samples against the variance of each update, and it is an integer multiple of the mini-batch size so that each collected batch is fully utilized during training.

To assess the benefit of the proposed PPO-based approach, we compare it with several baseline scheduling policies:
\begin{enumerate}
    \item [(i)] \textit{Deep deterministic policy gradient (DDPG)}~\cite{ddpg}, an off-policy algorithm that learns a deterministic policy by adapting Q-learning ideas to continuous action spaces.
    \item [(ii)] \textit{Twin delayed DDPG (TD3)}~\cite{td3}, which improves upon DDPG by employing clipped double Q-learning and delayed policy updates to mitigate function approximation errors and overestimation bias.
    \item [(iii)] \textit{Soft actor--critic (SAC)}~\cite{sac}, a maximum entropy RL framework that optimizes a stochastic policy to concurrently maximize the expected return and policy entropy for enhanced exploration and stability.
    \item [(iv)] \textit{A channel-gain-based greedy heuristic}, which serves as a performance lower bound for learning-based methods.
\end{enumerate}
In the greedy baseline, all device and satellite transmit powers are fixed at their maximum values, and, in each epoch, the $V$ cells with the strongest effective channel gains are selected for update.
This policy aggressively exploits the best instantaneous channels and, as a result, drives the aggregation MSE well below the prescribed threshold.
However, it does not explicitly trade off the MSE margin against the amount of collected training data.
All DRL-based baselines share the same state and action definitions and reward function as PPO, and are trained in the same environment with carefully tuned hyperparameters to ensure a fair comparison. 

\begin{table}[htbp]
\centering
\caption{Simulation Parameters}
\label{tab:sys-params}
\begin{tabular}{l l}
\hline
\textbf{Parameter} & \textbf{Value} \\
\hline
Orbit altitude & 550 km \\
Coverage radius of a cell & 30 km \\
Number of satellites $N$ & 6 \\
Maximum beams per satellite $V$ & 4 \\
Number of cells served per satellite $C$ & 16 \\
Number of cells covered by all satellites $S$ & 70 \\
Number of devices per cell & 3 \\
Bandwidth & 500 MHz \\
Carrier frequency & 20 GHz \\
Maximum device transmit power $P_k$ & 8.4 dBW \\
Maximum satellite transmit power $P_n$ & 30 dBW \\
Aperture radius of the antenna & 0.15 m \\
Satellite beam 3 dB beamwidth & 3$^{\circ}$ \\
Maximum device antenna gain $G_k^{\text{max}}$ & 0 dBi \\
Satellite antenna gain $G_m$ & 35.9 dBi \\
Noise temperature & 354.81 K \\
Epoch duration $\tau$ & 7.6 s \\
New data volume $|\bar{\mathcal{D}}_k(t)|$ & 30--50 samples/epoch \\
Maximum data buffer capacity $D_{\max}$ & 100 \\
Discount factor of data amount $\eta$ & 0.5 \\
Global MSE threshold $\rho$ & -5 dB \\
MSE penalty weight $\mu$ & 0.5 \\
\hline
\end{tabular}
\end{table}

\begin{table}[htbp]
\centering
\caption{Federated Learning Tasks and Model Parameters}
\label{tab:fl-params}
\begin{tabular}{l c c}
\hline
\textbf{Parameter} & \textbf{MNIST + MLP} & \textbf{CIFAR-10 + CNN} \\
\hline
Dataset size (train/test) & 60{,}000/10{,}000 & 50{,}000/10{,}000 \\
Model structure & [784, 512, 256, 10] & [2 Conv, 3 FC] \\
Detailed layers & -- & \makecell[l]{Conv: 6@5x5, 16@\\5x5 FC: 120, 84, 10} \\
Activation function & ReLU & ReLU \\
Local learning rate $\eta_l$ & 0.04 & 0.03 \\
Momentum & 0.5 & 0.9 \\
Local training epochs $E$ & 2 & 3 \\
Global rounds & 60 & 180 \\
\hline
\end{tabular}
\end{table}

\begin{table}[htbp]
\centering
\caption{DRL Training Parameters for Different FL Tasks}
\label{tab:drl-params}
\begin{tabular}{l c c}
\hline
\textbf{Parameter} & \textbf{MNIST + MLP} & \textbf{CIFAR-10 + CNN} \\
\hline
Training episodes & 1500 & 1500 \\
Actor network learning rate $\alpha_{\mathrm{a}}$ & 0.0005 & 0.001 \\
Critic network learning rate $\alpha_{\mathrm{c}}$ & 0.002 & 0.002 \\
Discount factor $\gamma$ & 0.95 & 0.95 \\
GAE parameter $\lambda_{\text{GAE}}$ & 0.95 & 0.95 \\
PPO clip ratio $\epsilon$ & 0.2 & 0.2 \\
Mini-batch size $B_{\text{PPO}}$ & 64 & 64 \\
Update interval $K_{\mathrm{PPO}}$ & 256 & 256 \\
Update epochs per iteration & 10 & 8 \\
Entropy coefficient $c_{\mathrm{e}}$ & 0.001 & 0.001 \\
Entropy coefficient decay $\delta_{c_{\mathrm{e}}}$ & 0.95 & 0.95 \\
Data type & IID & IID \\
\hline
\end{tabular}
\end{table}

\subsection{DRL Training Performance}

\begin{figure}[htbp]
\centering
\includegraphics[width=0.5\textwidth]{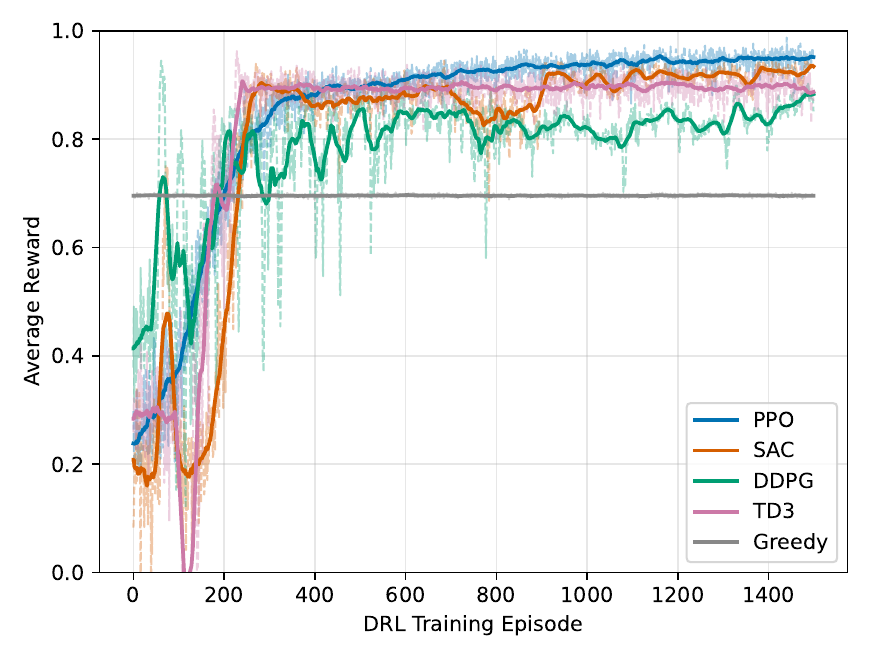}
\caption{Average training reward among different algorithms in the MNIST+MLP task.}
\label{fig:reward_mlp}
\end{figure}

\begin{figure}[htbp]
\centering
\includegraphics[width=0.5\textwidth]{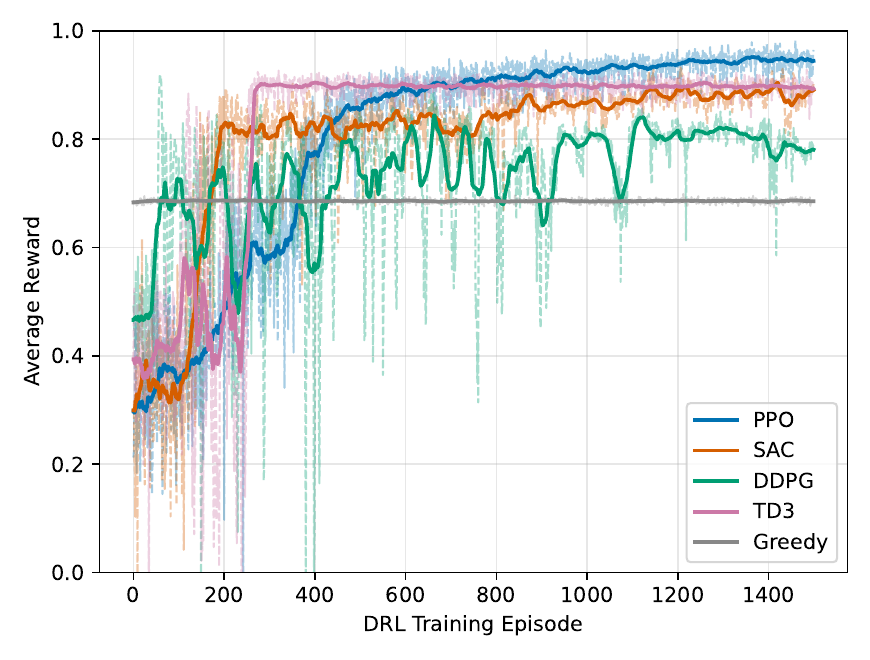}
\caption{Average training reward among different algorithms in the CIFAR-10+CNN task.}
\label{fig:reward_cnn}
\end{figure}

We first examine the learning behavior of the PPO-based algorithm. 
Fig.~\ref{fig:reward_mlp} and Fig.~\ref{fig:reward_cnn} depict the evolution of the average training reward for PPO, SAC, DDPG, TD3, and the greedy baseline in the MNIST+MLP and CIFAR-10+CNN tasks, respectively. 
In both figures, the solid curves represent the moving-average reward over a sliding window of 30 training episodes, while the faint dashed traces correspond to the raw episode-wise rewards. 

In the MNIST+MLP task (Fig.~\ref{fig:reward_mlp}), all DRL algorithms quickly improve their rewards during the first few hundred training rounds, and then gradually approach a steady regime. 
Among them, PPO exhibits a smooth and monotonic increase and attains the highest steady-state reward with relatively small fluctuations, indicating that it can reliably learn effective beam-hopping and power control policies under the global MSE constraint. 
SAC also converges to a high reward level but with slightly slower improvement and more noticeable oscillations, leading to a final performance that is consistently lower than that of PPO. 
TD3 achieves intermediate performance: it can rapidly reach a high reward region, but remains below PPO and SAC in the steady state. 
DDPG yields the lowest reward among the DRL methods and exhibits pronounced instability with recurrent dips, which reflects its sensitivity to the non-stationary dynamics of the OTA-enabled online FL environment. 
The greedy baseline, whose policy is fixed and does not learn over time, maintains an almost constant reward around 0.7, clearly lagging behind all DRL-based schedulers.

In the CIFAR-10+CNN task (Fig.~\ref{fig:reward_cnn}), the reward curves become more noisy due to the higher model dimensionality and more complex gradient statistics. 
Nevertheless, the overall trend is similar: PPO continues to improve throughout the training process and ultimately attains the highest reward among all schemes. 
In the early and intermediate stages, TD3 and SAC can momentarily reach relatively high reward values, but their improvements quickly saturate and their steady-state rewards remain noticeably lower than that of PPO. 
TD3 generally outperforms SAC in this more challenging task, while DDPG again exhibits the lowest reward and the strongest oscillations. 
The greedy baseline stays nearly flat and is consistently inferior to any learning-based policy. 
These results demonstrate that the proposed PPO-based approach not only converges more stably than the other DRL methods, but also generalizes well across FL tasks with substantially different complexity levels.

\begin{figure}[htbp]
\centering
\includegraphics[width=0.5\textwidth]{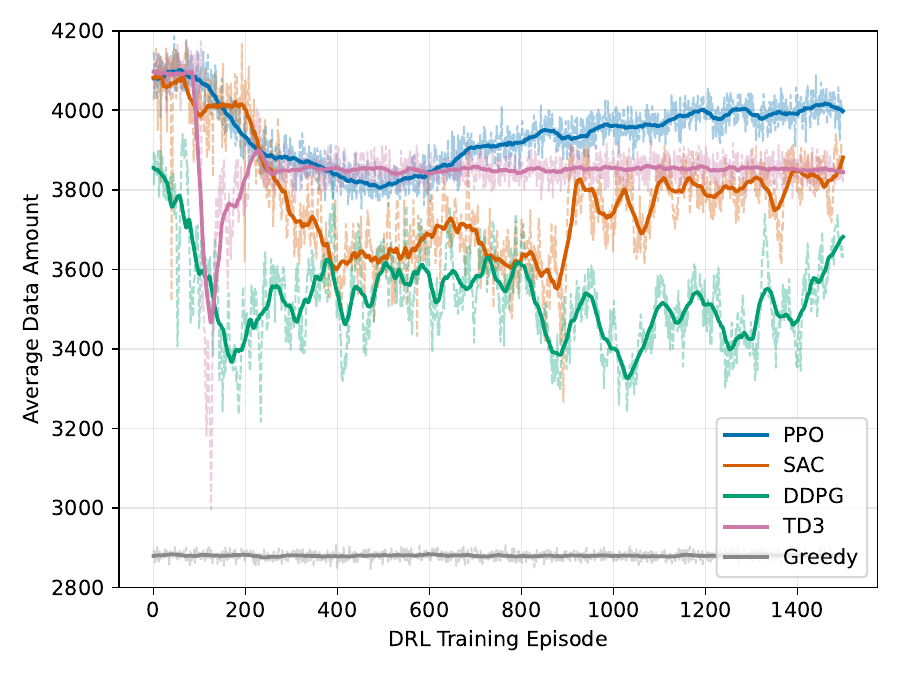}
\caption{Average data amount versus training round among different algorithms in the MNIST+MLP task.}
\label{fig:dataamount_episode}
\end{figure}

To further understand how different schedulers exploit the available training data over time, Fig.~\ref{fig:dataamount_episode} depicts the evolution of the data amount versus the training round in the MNIST+MLP scenario for all considered algorithms. 
As in Figs.~3 and~4, the solid curves are smoothed using the same moving average over a sliding window of $30$ training episodes, while the faint traces show the raw per-episode values.
At the early stage of training, all DRL-based policies experience an apparent drop in data amount. 
This phenomenon is mainly because the initial policies are not yet MSE-aware and frequently violate the prescribed MSE threshold, incurring the penalty term in the reward and thus driving the learned behaviors toward more conservative actions to avoid MSE violations. 
As training proceeds, the agents gradually learn to satisfy the MSE requirement while increasing data utilization, leading to a sustained rise of the data-amount curves. 
Among them, the curve of PPO keeps increasing and eventually stabilizes at the highest level, indicating that the learned policy can consistently collect more usable data per round. 
SAC and TD3 converge to intermediate data levels with moderate variability, while DDPG remains lower and shows stronger oscillations, reflecting its less reliable behavior in the non-stationary environment. 
In contrast, the greedy baseline stays almost flat at a relatively low data amount throughout training, as it repeatedly schedules only a limited set of well-conditioned cells. 
Overall, together with the reward trends in Fig.~\ref{fig:reward_mlp} and Fig.~\ref{fig:reward_cnn}, these results confirm that PPO achieves superior long-term data utilization by learning an MSE-compliant scheduling policy.

\subsection{FL Performance on MNIST and CIFAR-10}

\begin{figure}[htbp]
\centering
\includegraphics[width=0.5\textwidth]{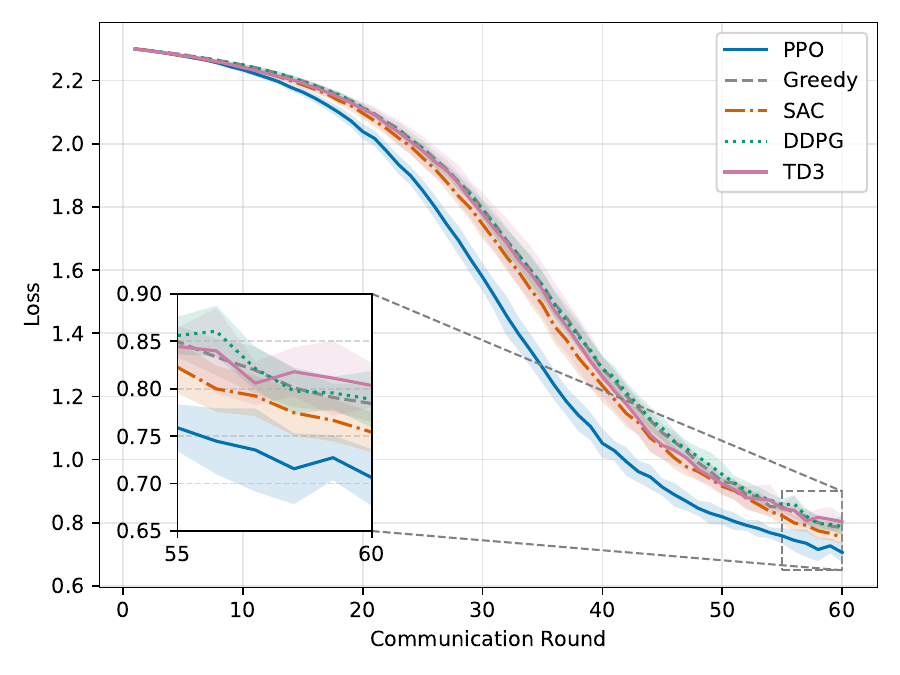}
\caption{Global training loss versus communication round for the MNIST+MLP task among different algorithms.}
\label{fig:mnist_loss}
\end{figure}

\begin{figure}[htbp]
\centering
\includegraphics[width=0.5\textwidth]{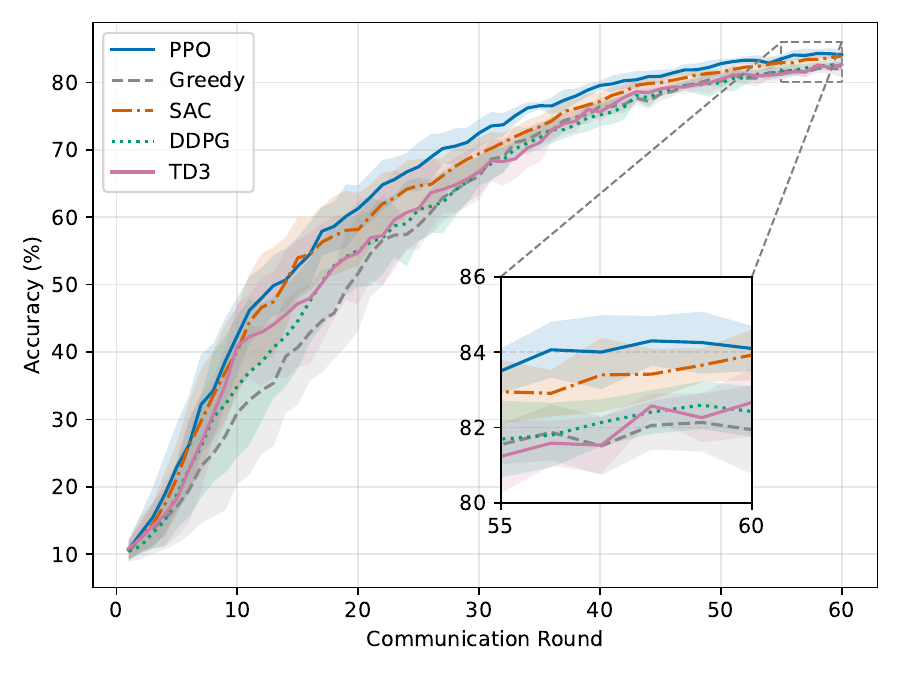}
\caption{Global test accuracy versus communication round for the MNIST+MLP task among different algorithms.}
\label{fig:mnist_acc}
\end{figure}

We next evaluate how the learned scheduling policies affect the FL convergence behavior. 
Fig.~\ref{fig:mnist_loss} and Fig.~\ref{fig:mnist_acc} show the global training loss and test accuracy, respectively, as functions of the communication round for the MNIST+MLP task. 
Each curve reports the average over 5 independent runs with different random seeds, and the shaded bands indicate the corresponding standard deviation. 
Under all schemes, the loss decreases and the accuracy increases monotonically with the communication round, confirming that the OTA aggregation with the imposed MSE constraint enables effective federated training. 
However, noticeable differences appear among the scheduling policies. 
PPO achieves the fastest loss reduction, especially between rounds 20 and 40, and attains the lowest final loss after 60 rounds. 
The corresponding accuracy curve clearly dominates the baselines over almost the entire training process: PPO reaches around 70\% accuracy by roughly round~25, and finally converges to above 84\%. 
SAC and TD3 form a second tier; they exhibit similar convergence speeds and eventually approach PPO, but remain consistently below it in both loss and accuracy. 
DDPG and the greedy baseline converge more slowly and yield slightly higher final loss and lower final accuracy, with the greedy policy performing particularly poorly in the early rounds due to its tendency to repeatedly schedule a small set of well-conditioned cells. 
These results indicate that the PPO-based scheduler provides a better accuracy–round trade-off on the relatively light-weight MNIST task.

\begin{figure}[htbp]
\centering
\includegraphics[width=0.5\textwidth]{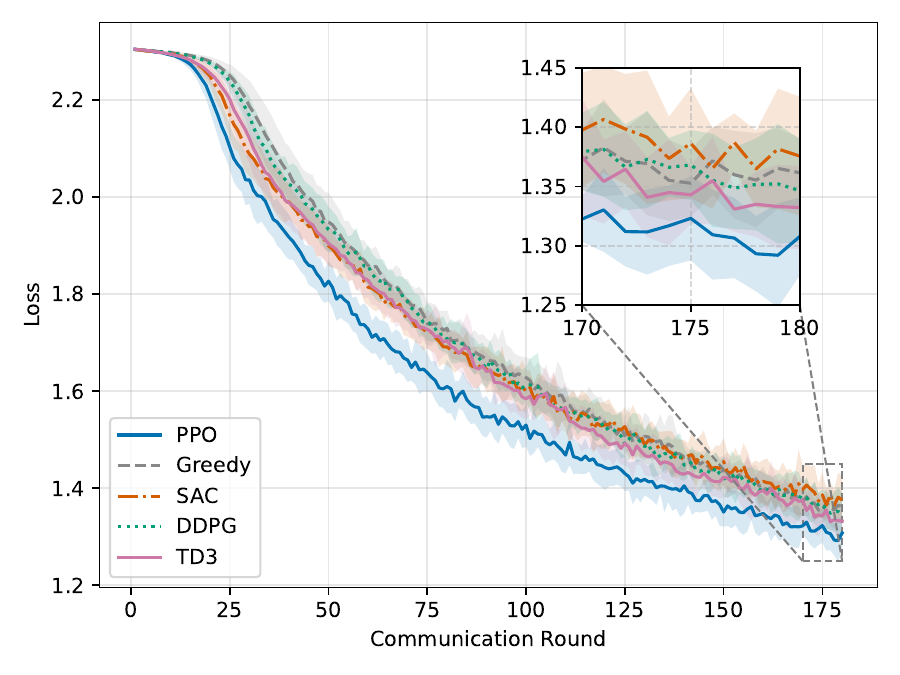}
\caption{Global training loss versus communication round for the CIFAR-10+CNN task.}
\label{fig:cifar_loss}
\end{figure}

\begin{figure}[htbp]
\centering
\includegraphics[width=0.5\textwidth]{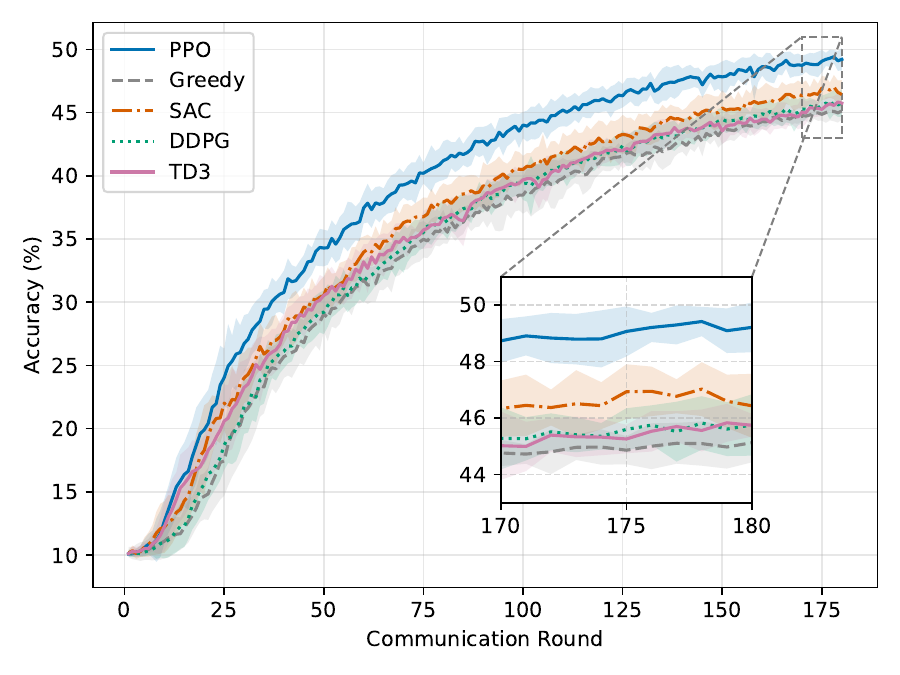}
\caption{Global test accuracy versus communication round for the CIFAR-10+CNN task.}
\label{fig:cifar_acc}
\end{figure}

For the more challenging CIFAR-10+CNN task, the evolution of the global loss and test accuracy is depicted in Fig.~\ref{fig:cifar_loss} and Fig.~\ref{fig:cifar_acc}, respectively. 
Compared with the MNIST case, all methods exhibit higher residual loss and lower final accuracy, which is expected given the higher model dimensionality and more complex data distribution. 
Nevertheless, the performance ranking of the scheduling policies is largely preserved. 
PPO consistently attains the lowest loss throughout training and maintains a visible margin over all baselines. 
In terms of accuracy, PPO again dominates the entire trajectory, achieving the fastest initial growth and the highest final accuracy. 
SAC and TD3 follow as the next-best performers, while DDPG and the greedy baseline lag behind with noticeably higher loss and lower accuracy. 
The gap between PPO and the greedy policy is more pronounced on CIFAR-10 than on MNIST, highlighting that blindly minimizing the aggregation MSE (as greedy does) is insufficient when the FL workload becomes more complex, and that intelligently trading off MSE margin against long-term data utilization is crucial for maintaining good learning performance. 
Overall, the MNIST and CIFAR-10 results together demonstrate that the proposed PPO-based scheduler consistently improves FL convergence speed and final model quality across tasks with substantially different difficulty levels.

\subsection{Impact of BH and MSE Threshold on Data Utilization}

\begin{figure}[htbp]
\centering
\includegraphics[width=0.5\textwidth]{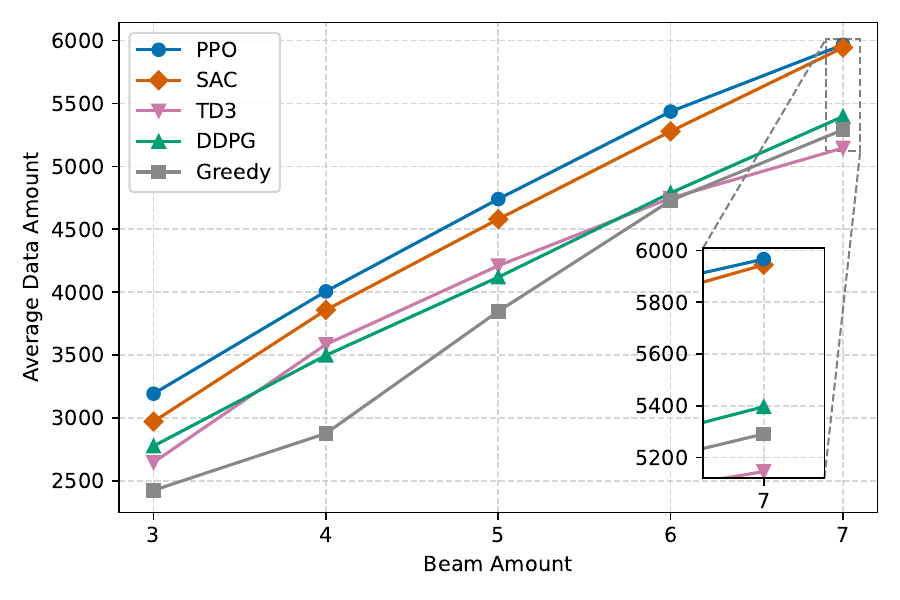}
\caption{Average data amount per round versus the maximum number of beams $V$ per satellite among different algorithms.}
\label{fig:beam_dataamount}
\end{figure}

\begin{figure}[htbp]
\centering
\includegraphics[width=0.5\textwidth]{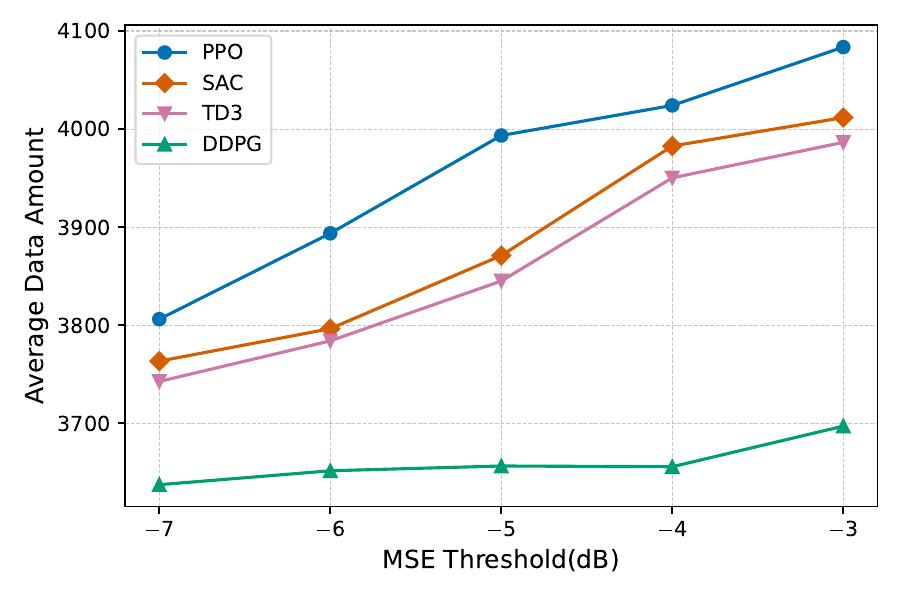}
\caption{Average data amount per round versus the global MSE threshold $\rho$ among different algorithms.}
\label{fig:mse_dataamount}
\end{figure}

We finally study how key system parameters influence the amount of usable training data collected by different scheduling policies. 
Fig.~\ref{fig:beam_dataamount} reports the average data amount per round as a function of the maximum number of beams $V$ per satellite, while Fig.~\ref{fig:mse_dataamount} shows the impact of the global MSE threshold~$\rho$. 
Both figures are obtained under the MNIST+MLP task setting, and the plotted data amounts are averaged over 60 FL communication rounds.

As shown in Fig.~\ref{fig:beam_dataamount}, increasing the beam budget $V$ leads to a nearly linear growth in the average data amount for all schemes, since more beams enable each satellite to activate more cells and hence aggregate more device updates per communication round. 
Throughout the range from $V=$ 3 to $V=$ 7, PPO consistently achieves the highest data utilization.
While SAC, TD3, and DDPG also benefit from larger $V$, their gains are comparatively smaller, and the greedy heuristic remains the worst-performing approach. 
Overall, these results suggest that PPO provides robust data-utilization improvement over a wide range of beam budgets, and the advantage persists even in the high-beam regime.

Fig.~\ref{fig:mse_dataamount} illustrates how the global MSE threshold $\rho$ affects the data utilization of the DRL-based schedulers. 
Relaxing the threshold from -7~dB to -3~dB monotonically increases the average data amount for all algorithms, as a looser distortion requirement allows the system to aggregate more noisy updates. 
However, the rate of increase and the absolute performance significantly differ across policies. 
PPO yields the highest data amount under all considered thresholds and shows the most pronounced improvement when~$\rho$ becomes less stringent, demonstrating its ability to exploit the additional MSE margin while still controlling the long-term penalty in the reward. 
SAC and TD3 form a second tier, lagging behind PPO but clearly outperforming DDPG, whose data amount remains almost flat for most thresholds and only slightly increases at $\rho=$-3~dB. 
Combined with the FL accuracy and loss results in Figs.~\ref{fig:mnist_loss}--\ref{fig:cifar_acc}, these observations confirm that the PPO-based approach can effectively adapt to different beam and MSE configurations and consistently extract more useful training data from the underlying satellite network.

\section{Conclusion}

This paper investigated adaptive BH and power control for OTA-enabled online FL in LEO satellite networks, where ground devices continuously collect new data and untrained samples gradually lose freshness. We proposed a dual-layer OTA aggregation framework that connects devices, satellites, and a DPC, and formulated a long-term optimization problem that maximizes the discounted amount of utilized training data under satellite-specific beam, power, and global MSE constraints. The problem was cast as an MDP and tackled by a PPO-based DRL algorithm that jointly learns beam-hopping patterns and power-control strategies from interaction with the environment. Extensive simulations on MNIST+MLP and CIFAR-10+CNN tasks showed that the proposed PPO-based algorithm consistently outperforms SAC, DDPG, TD3, and a channel-gain-based greedy policy in terms of training reward, final training loss and test accuracy, and long-term data utilization. 
Sensitivity results further indicate that increasing the beam budget and a less stringent MSE threshold both improve data utilization, while the proposed PPO-based scheduler consistently achieves the highest data amount by leveraging the available degrees of freedom more effectively.
We note that the present study relies on two simplifying assumptions that also delineate its main limitations. First, the local data across devices are assumed to be independent and identically distributed (IID), whereas real-world satellite IoT deployments may exhibit significant statistical heterogeneity across geographically dispersed regions. Second, the channel model assumes perfect CSI at the scheduler, and imperfect or delayed CSI would degrade the channel-inversion alignment and the OTA aggregation accuracy.
Future work will extend the framework to multi-satellite cooperative aggregation with time-varying network topologies and model heterogeneity, and investigate theoretical performance guarantees under delayed and imperfect channel state information.
\bibliographystyle{IEEEtran}
\bibliography{reference}
 
\end{document}